\documentstyle[preprint,prl,aps,psfig]{revtex}
\tightenlines
\begin{document}
\draft
\title{Quantum dots based on spin properties of semiconductor
heterostructures}
\author{Manuel Val\'{\i}n-Rodr\'{\i}guez$^1$, Antonio Puente$^1$
and Lloren\c{c} Serra$^{1,2}$}
\address{
$^1$ Departament de F\'{\i}sica, Universitat de les Illes Balears,
E-07122 Palma de Mallorca, Spain\\*
$^2$ Institut Mediterrani d'Estudis Avan\c{c}ats IMEDEA (CSIC-UIB),
E-07122 Palma de Mallorca, Spain}
\date{October 5, 2003}
\maketitle
\begin{abstract}

The possibility of a type of semiconductor 
quantum dots obtained by spatially modulating the spin-orbit 
coupling intensity in III-V heterostructures is discussed.
Using the effective mass model we predict confined one-electron states 
having peculiar spin properties.
Furthermore, from mean field calculations (local-spin-density and Hartree-Fock)
we find that even two electrons could form a bound state
in these dots.
 
\end{abstract}

Low-dimensional semiconductor structures constitute one of the
most suitable scenarios to observe quantum phenomena in condensed 
matter physics. In this sense, the atomic-like properties of quantum
dots and the conductance quantization in narrow wires are
seminal examples widely studied \cite{haw97,datta97}. 
At present, the available fabrication techniques of such
semiconductor structures rely on charge-based confinement at
length scales where quantum effects are unavoidable. More precisely, 
confinement is obtained through the energy band discontinuities of 
different materials or by directly creating a potential landscape 
by means of metallic electrodes on a semiconductor heterostructure.

The aim of this work is to point out, using a theoretical model, the 
feasibility of a class of mesoscopic semiconductor structures 
where the spin plays a relevant role in the mechanism that determines 
the confining properties, without being a pure spin-dependent confinement 
\cite{fleu01}. The mechanism also differs from conventional electrostatic 
confinement in that it does not rely on the direct coupling of the electron
charge with a horizontal electric field.  

In III-V non-magnetic semiconductor heterostructures there are two
relevant interactions concerning the spin: the spin-orbit coupling 
and the hyperfine interaction between carriers and polarized nuclei.
The latter mechanism has a characteristic energy scale usually  
lower than that of the spin-orbit coupling \cite{fleu01}.

In semiconductors, spin-orbit coupling stems from the relativistic
effect caused by the electric field due to the lack of 
inversion symmetry of certain alloys like III-V heterostructures.
Depending on the particular origin of the electric field two
contributions are distinguished; the electric field created by the
bulk inversion asymmetry (BIA) of the material gives rise to the
Dresselhaus term \cite{dres55}, while the asymmetry in the profile of the 
heterostructure, structural asymmetry (SIA), generates the so-called
Bychkov-Rashba term \cite{ras60}. This latter mechanism constitutes the 
basis of the proposed electronic nanostructures.

As mentioned above, the intensity of the Bychkov-Rashba spin-orbit coupling
depends on the effective electric field perpendicular to the plane of the quantum
well. This implies that it can be, in principle, engineered by adjusting the 
specific fabrication settings of each heterostructure. Another possibility
to control the Rashba interaction, that has been  
demonstrated experimentally \cite{nitt97,can99,gru00},
is by means of an external electric field perpendicular
to the plane of the quantum well.

Exploiting the tunability of the Bychkov-Rashba mechanism we consider 
quantum dots whose main ingredient is a space-modulated spin-orbit
intensity. This spatial modulation could be 
obtained with biased small electrodes on top of the semiconductor
heterostructure\cite{halp02,val03}. However, such an arrangement would not 
only modify the vertical electric field but it would also generate
an in-plane gradient of electric potential, leading to hybrid quantum dots
where both mechanisms are present: horizontal electric potential gradient 
and spin-orbit intensity gradient.
A more refined method to obtain space modulated spin-orbit intensities
requires a deeper insight in the origin of the Rashba intensity.
In this respect, de Andrada {\em et al} \cite{and97} discussed the relevance 
of the barrier layers in the determination of the spin-orbit intensity. 
More recently, Grundler \cite{gru00} has shown that 
for an InAs-based quantum well an applied vertical electric field of the 
order of $10^3$ V/cm can lead
to a noticeable control of the spin-orbit intensity, whereas the built-in electric
field of the sample is of the order of $10^5$ V/cm. 
This high sensitivity to a relatively weak electric
field has been attributed to the induced variations in the wavefunction 
penetration into the barriers.
The strong relevance of the barrier layers in the determination of the Rashba
intensity supposes a great advantage that could help to create a quantum well 
having a space-modulated spin-orbit coupling through space-engineered
wavefunction barrier penetration
(using inhomegeneous barrier layer composition, thickness, ...). 
This would lead to a fixed (not tunable) substantial step in the spin-orbit 
intensity capable of inducing the confinement.

We model structures confining electrons from the conduction band
of the semiconductor and we use the effective mass
approximation. We also suppose that the motion perpendicular to the
well's plane is restricted to the first transversal subband, so that 
the electrons effectively move in a two-dimensional region. These 
simplifications might be quite drastic in some cases and, therefore, 
the scope of this work 
is limited to a discussion of the feasibility of the effects rather 
than to a detailed quantitative description.  
In this model we include the standard Bychkov-Rashba term 
\begin{eqnarray}
\label{eq1}
{\cal H}_{R} = \frac{\lambda_R(r)}{\hbar} 
\left(\, p_y\sigma_x-p_x\sigma_y\,\right) \; ,
\end{eqnarray}
where $p_x$ and $p_y$ represent the components of the in-plane electron's
momentum and the $\sigma$'s are the corresponding Pauli matrices. 

In Eq.\ (\ref{eq1}) the inhomogeneity of the spin-orbit intensity is 
included through the spatial dependence of the coupling constant 
$\lambda_R(r)$, which is assumed of radial type ($r$ and $\theta$ 
are used to label the standard polar coordinates in the plane). 
To characterize the quantum dot we consider the 
following step-like variation of the spin-orbit intensity
\begin{equation}
\lambda_R(r)=\lambda_e+(\lambda_i-\lambda_e)\frac{1}{1+e^{\frac{r-R_0}{\sigma}}}
\; ,
\end{equation}
where $\lambda_e$ and $\lambda_i$ represent the external 
and internal constant values of spin-orbit intensity, respectively.
The quantum dot radius is given by $R_0$ and $\sigma$ is 
a small diffusivity introduced to avoid discontinuities in the numerical 
solutions.  

The Bychkov-Rashba interaction breaks the rotational
symmetry of the Hamiltonian. Nevertheless, a symmetry combining spatial
and spin degrees of freedom is maintained when the Rashba intensity is
a purely radial function. Namely, the Hamiltonian is invariant under a combined
rotation in real and spin spaces by the same angle, i.e., it 
commutes with the generalized angular momentum $J_z = L_z + S_z$. 
Note that, at variance with conventional circular quantum dots
(based on electric-potential confinement \cite{haw97}),
the Hamiltonian no longer commutes with $L_z$ and $S_z$ separately.
As a consequence of the $J_z$-symmetry the single-particle solutions 
can be expressed, in a general form, as two-component spinors of type
\begin{equation}
\label{eq3}
 {\bf  \Psi}_{nj}(r,\theta) = 
 \left(\begin{array}{c}
	\phi_{nj\uparrow}(r)e^{i(j-1/2)\theta} \\
	\phi_{nj\downarrow}(r)e^{i(j+1/2)\theta}
	\end{array}\right)\; ,
\end{equation}
where $n$ and $j$ are quantum numbers characterizing the states,
with values $n=1,2,\dots$, $j=\pm1/2,\pm3/2,\dots$.
Since the angular parts are analytical, we can reduce the Schr\"odinger 
equation to one-dimensional radial equations for the two spinor components,
\begin{eqnarray}
-\frac{\hbar^2}{2m^*}
\left[
\frac{\partial\phi_\uparrow}{\partial r^2}+ 
\frac{1}{r} \frac{\partial \phi_\uparrow}{\partial r}-
\frac{(j-1/2)^2}{r^2}
\phi_{\uparrow}
\right]
-\lambda_{R}(r)\left[
\frac{\partial \phi_\downarrow}{\partial r} + 
\frac{j+1/2}{r} \phi_{\downarrow}\right]
= \varepsilon_{nj}\, \phi_{\uparrow}\nonumber\\
-\frac{\hbar^2}{2m^*}
\left[
\frac{\partial\phi_\downarrow}{\partial r^2}+ 
\frac{1}{r}
\frac{\partial\phi_\downarrow}{\partial r}-\frac{(j+1/2)^2}{r^2}
\phi_\downarrow \right]
+\lambda_{R}(r)\left[
\frac{\partial \phi_\uparrow}{\partial r} 
- \frac{j-1/2}{r} \phi_{\uparrow}\right]
=\varepsilon_{nj}\, \phi_{\downarrow}
\label{eq4}
\end{eqnarray}
where $\phi_{\uparrow}$ and $\phi_{\downarrow}$ denote $\phi_{nj\uparrow}(r)$ 
and $\phi_{nj\downarrow}(r)$, respectively.

The above model has been analyzed using two alternative numerical procedures: 
a) solving the coupled one-dimensional
equations for the radial components of the eigenspinors, Eqs.\ (\ref{eq4}); and
b) solving the full two-dimensional problem in Cartesian coordinates, 
without imposing any symmetry restriction, in a uniform grid.
Before discussing the numerical results it has to be noted that systems with the
above spin-orbit interaction are still invariant under time reversal. 
Consequently, a two-fold degeneracy, known as Kramers degeneracy, must hold. 
From the radial equations it can easily be seen that the substitutions: 
$\phi_{\uparrow}\rightarrow\phi_{\downarrow}$, $\phi_{\downarrow}\rightarrow-\phi_{\uparrow}$
and $j\rightarrow -j$ lead to a new state (Kramers  
conjugate) with opposite generalized angular momentum and 
the same energy eigenvalue ($\varepsilon_{nj}=\varepsilon_{n-j}$). 
The expectation values $\langle S_z\rangle$ and $\langle L_z\rangle$
also change sign with the Kramers transformation. 
It is worth stressing that in the usual quantum dots (with an 
electric potential confinement of circular symmetry) the degeneracies
are 4 and 2; nonzero (zero) angular momentum states being four-fold (two-fold) 
degenerate. Therefore, the reduction in degeneracy from 4 to at most 2 
is a genuine effect of the spin-orbit interaction.

Since the spin orientation in Eq.\ (\ref{eq3}) depends
on the position ${\bf r}$, the eigenstates will show characteristic
spin textures. The distribution of vertical
and parallel spin read
\begin{eqnarray}
\langle{\sigma}_z\rangle(r) &=& 
|\phi_{nj\uparrow}(r)|^2-|\phi_{nj\downarrow}(r)|^2
\nonumber\\
\langle\vec{\sigma}_\parallel\rangle(r,\theta) &=& 
2\phi_{nj\downarrow}(r)\phi_{nj\uparrow}(r)
\left[\cos(\theta)\hat{x}+\sin(\theta)\hat{y}\right]\; .  
\end{eqnarray}
Note that while the vertical component is a pure radial function,
the in-plane (horizontal) spin points in the radial direction. The
fact that the horizontal spin follows the position vector can be easily 
explained from the combined rotational invariance in spin and position
space: consider two points in the dot having the same radius and different 
azimuthal angle $\theta$;
due to the symmetry of the Hamiltonian with $J_z$, the local in-plane spin 
orientation in the second point must be that of the first point rotated the
angle needed to go from the first point to the second in the real space.
This property of the eigenstates is of particular interest for determining the role
of the geometrical phase (Berry's phase) in mesoscopic structures 
\cite{vag98,aro93}.

From the numerical simulations we observe that a necessary requirement for 
confined states is that the spin-orbit intensity
within the quantum dot must be larger 
than that of the outer region, i.e., $\lambda_i>\lambda_e$.
The reason for this behaviour can be understood from bulk considerations: 
the bulk energy bands
\begin{eqnarray}
\label{off}
\varepsilon^{\rm bulk}_{k\pm} &=& {\hbar^2 \over 2m^*}\left(k\pm{m^*\over\hbar^2}\lambda_e\right)^2
-{m^*\over 2\hbar^2}\lambda_e^2\; 
\end{eqnarray} 
display a constant offset $-m^*\lambda_e^2/2\hbar^2$ 
that depends on the spin-orbit intensity
(for a schematic representation of the bulk bands see Fig.\ 1).
Therefore a bulk uniform sample with a big value of
spin-orbit intensity has an energy origin for the eigenstates lower than that of
a sample having a smaller intensity. In a finite system this energy mismatch 
can lead to bound states in the region of higher intensity. Nevertheless, it has to
be noted that for each spin orientation the energy mismatch is only confining for a 
certain range of momenta. This means that the confining mechanism is spin-selective
through the motion of the particles. In the case of circular symmetry, it implies 
that electrons rotating in one sense are confined if they have a given spin 
orientation, while electrons rotating in the opposite sense should have the opposite
spin orientation to be confined.

We stress that the SO-dependent offset in Eq.\ (\ref{off}) is the 
real cause of confinement since this term acts as an electrostatic 
potential when the SO intensity is varying in 
space. Actually, a different SO Hamiltonian in which this offset were not included
would not lead to confined states.

Another question of relevance refers to the magnitude of the step in the spin-orbit
intensity needed to obtain confined states. For a typical size
of dot radius of the order of 10 atomic effective units\cite{uni00} 
simulation shows that an intensity in the range $\lambda_i\simeq 1.5-2.0\times 10^{-9}$
eVcm is required to obtain a single-particle bound state when the external intensity
is zero ($\lambda_e=0$). 
It is also important to specify the external value $\lambda_e$, since the minimum step 
required to achieve confined states strongly depends on this quantity. 
For instance, when the external
intensity takes a value of the order of $\lambda_e \simeq 2\times 10^{-9}$ eVcm 
the step needed is reduced and the required internal value is in the range
$ \lambda_i -\lambda_e  \simeq 0.4-0.7\times 10^{-9}$ eVcm.

Quantum dots satisfying the required conditions stated above show a level structure
analogous to that of conventional quantum dots. For instance, a dot characterized
by spin-orbit parameters in the range of the biggest Rashba intensities experimentally
measured \cite{gru00}, $\lambda_i\simeq 4\times 10^{-9}$ eVcm and 
$\lambda_e\simeq 2\times 10^{-9}$ eVcm,
shows a total of 14 single-particle bound states distributed in two subbands (see Fig.\ 1).
The first one containing 10 states characterized by the quantum numbers 
$n=1$, $j=-9/2,-7/2,\dots,7/2,9/2$ and the second one having the remaining 4 states 
$n=2$, $j=-3/2,-1/2,1/2,3/2$. 
Figure 2 shows the spatial representation of the lowest energy state of this dot 
corresponding to $n=1, j=1/2$ and its conjugated state $n=1, j=-1/2$. The upper panels 
show the radial profile of the spinorial components for both states, the only
differences between the two cases are the exchange of the spin index and
the sign change in one of the components, in agreement with symmetry properties
derived above. Middle panel shows the spatial distribution of the density which
has radial symmetry and is common for both conjugated states. Finally, the lower
panel displays the radial distribution of the $S_z$ spin density for the two states,
showing that conjugated states posses opposite spin characters.

Up to now, we have only presented results concerning the single-particle 
properties of the dots. There is also the possibility of the spin-orbit-based
confinement to be robust against electron-electron interaction and, thus,
able to support multielectron bound states. To treat the 
interacting problem we impose no symmetry restriction
in space and we solve numerically the full two-dimensional problem including 
the interaction within density functional theory, using a generalization 
of the local-spin-density approximation (LSDA) for non-collinear spin 
densities (see Ref.\ \onlinecite{val01} for a previous application of this
formalism).
Within this scheme, we find that even two-electron states can be 
confined to the dot. Figure 3 shows the ground state density for an interacting
two-electron dot characterized by a radius of 15 units and the same spin-orbit
parameters of Fig.\ 2. The density displays a circular ring-like shape  
although the spin-orbit term breaks the rotational invariance. As an 
additional check of this bound two-electron state, we have also solved 
the Hartree-Fock equations, finding a similar total energy and 
density distributions also confined to the dot region. Extending the
calculations to treat three or more interacting electrons 
with the same spin-orbit term does not lead to confined states.  

In summary, we have proposed a mechanism to create electronic spatial 
confinement based on the spin properties of III-V semiconductor structures 
and, more specifically, on the Bychkov-Rashba spin-orbit interaction and 
its strong dependence on the barrier layer composition.
The eigenstates confined to these dots show common characteristics with
conventional semiconductor quantum dots such as a discrete level structure 
of the single-particle states and the possibility of confined two-electron  
states. The peculiar origin of the confinement reflects in the
spin properties of the single-particle eigenstates.
Not any material is suitable to create this type of structures since, 
as stated above, Rashba spin-orbit intensities of the order $10^{-9}$
eVcm and a range of tunability of the same magnitude are required. In this
sense, InGaAs-based wells \cite{can99} could be enough to produce the confining 
effect at mesoscopic scale. However, since the bulk-intensity value $\lambda_e$ 
has a strong relevance in the determination of the minimum conditions for confinement, 
InAs-based wells reveal as the best candidates to fabricate
these dots. This is because of their large built-in Rashba intensities, 
as large as $2\times 10^{-9}$ eVcm, and their tunability range of the same 
order\cite{gru00}.

This work was supported by Grant No.\ BFM2002-03241 
from DGI (Spain).

\begin{figure}[f]
\centerline{\psfig{figure=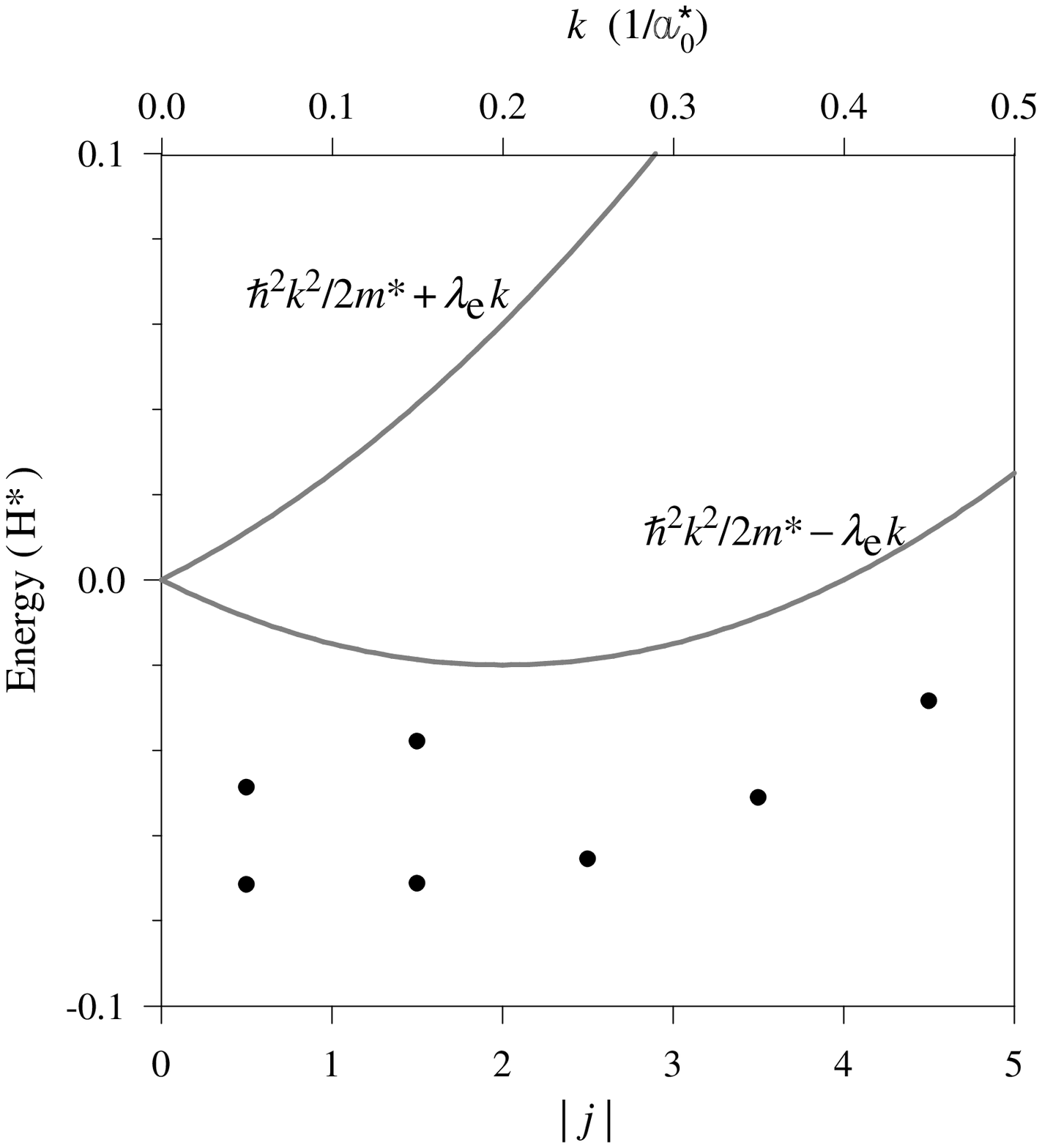,width=2.75in,clip=}}
\caption{Structure of the bound states as a function
of the generalized angular momentum $j$ for a quantum dot having  
$\lambda_e = 0.2\, H^*a_0^*\simeq 2\times 10^{-9}$ eVcm, 
$\lambda_i =0.4\, H^*a_0^*\simeq 4\times 10^{-9}$ eVcm and a nominal dot 
radius $R_0=10$ effective atomic units (lower scale). Solid lines represent  
the bulk bands corresponding to the surrounding material characterized
by $\lambda_e$ (upper scale).
}
\end{figure}

\begin{figure}[f]
\centerline{\psfig{figure=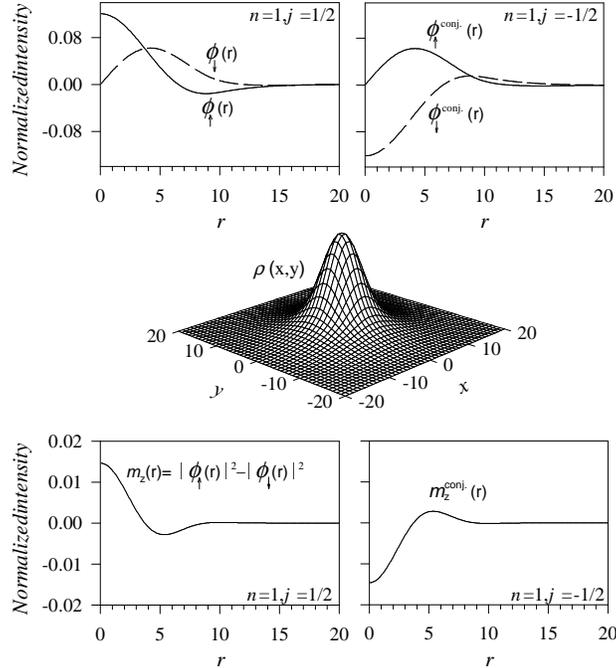,width=3.2in,clip=}}
\caption{
Upper panels: radial spinorial components corresponding
to the lowest energy pair of conjugated states.
Middle panel: two-dimensional representation of the probability
density corresponding to the above Kramers conjugated states. 
Lower panels: radial profile of
the $\sigma_z$ spin density distribution corresponding to the
same conjugated states. All coordinates are expressed in
terms of atomic effective units of distance 
}
\end{figure}

\begin{figure}[f]
\centerline{\psfig{figure=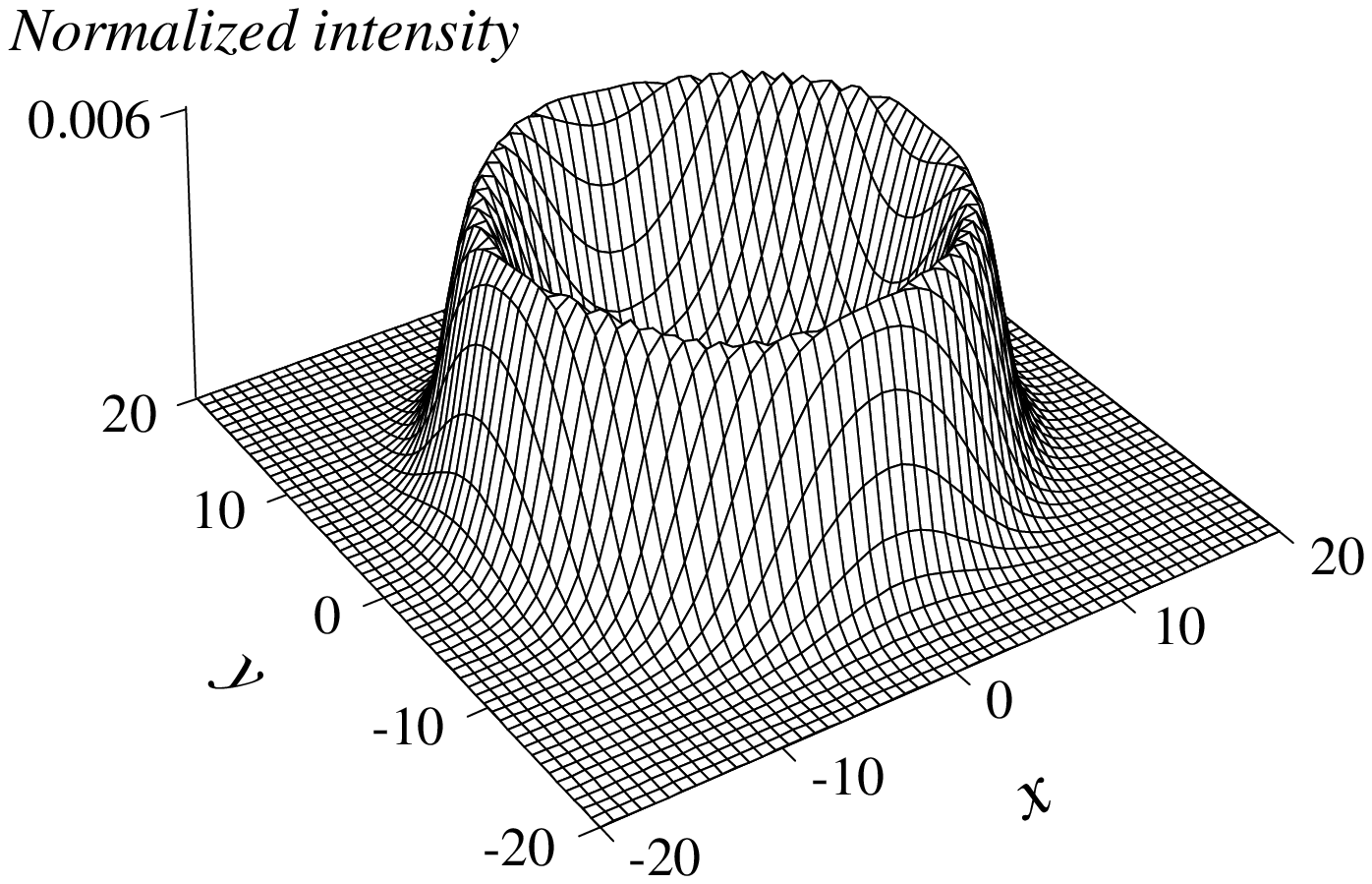,width=2.75in,clip=}}
\caption{Two-dimensional representation of the LSDA density 
corresponding to an interacting two-electron dot characterized
by $\lambda_e= 0.2\, H^*a_0^*\simeq 2\times 10^{-9}$ eVcm, 
$\lambda_i= 0.4\, H^*a_0^*\simeq 4\times 10^{-9}$ eVcm and a dot 
radius of 15 units. Atomic effective units of distance are used to 
represent the Cartesian coordinates.}
\end{figure}


\begin{thebibliography}{99}
\bibitem{uni00}
Atomic effective units take values in the range of 10-50 nm for distance ($a_0^*$)
and 5-10 meV for energy ($H^*$),  using typical semiconductor parameters,
while atomic effective units of spin-orbit intensity are more homogeneus
for the different materials having a good approximate value of
$1\times 10^{-8}$ eVcm.
\end{thebibliography}
\end{document}